\documentstyle[psfig,12pt,aaspp4]{article}
\begin{document}

\def\kms{{\rm km/s}}
\def\cc{{\rm cm ^{-3}}}
\def\mum{\mu {\rm m}}
\def\kpc{{\rm kpc}}
\def\deg{^\circ}
\def\hpc{{\rm h}^{-1}\,{\rm pc}}
\def\lsun{{\,L_\odot}}
\def\g0{{\,{\rm G}_0}}
\def\lcii{{\rm F_{\rm [CII]}}}
\def\l2cii{{\rm L_{\rm [CII]}}}
\def\ffir{{\rm F_{\rm FIR}}}
\def\lfir{{\rm L_{\rm FIR}}}
\def\rat{{\rm L_{\rm [CII]}}/{\rm L_{\rm FIR}}}
\def\2rat{{\rm L_{\rm [CII]}}/{\rm L_{\rm TIR}}}
\def\orat{{\rm L_{\rm [OI]}}/{\rm L_{\rm FIR}}}
\def\fb{{\rm F_{\rm B}}}
\def\7f{{\rm F_{7 \mu m}}}
\def\lb{{\rm L_{\rm B}}}
\def\6f{{\rm F_\nu(60 \mum)}}
\def\f100{{\rm F_\nu(100 \mum)}}
\def\etal{et al.~}
\def\HI{\ion{H}{1}}
\def\HII{\ion{H}{2}}
\def\IRAS{{\it IRAS}}
\def\ISO{{\it ISO}}
\def\wig#1{\mathrel{\hbox{\hbox to 0pt{%
  \lower.5ex\hbox{$\sim$}\hss}\raise.4ex\hbox{$#1$}}}}

\title{Evidence for the Heating of Atomic Interstellar Gas by PAHs}

\author{George Helou, \altaffilmark{1} 
Sangeeta Malhotra, \altaffilmark{2,3}
David J. Hollenbach, \altaffilmark{4}
Daniel A. Dale, \altaffilmark{1}
Alessandra Contursi \altaffilmark{1}}

\begin{abstract}

We report a strong correlation between the [CII] 158 $\mum$ cooling line
and the mid-infrared flux in the 5-10 $\mum$ range in a wide variety of
star-forming galaxies. The mid-infrared flux is dominated by Aromatic
Feature Emission (AFE), which is thought to arise from large polycyclic
aromatic hydrocarbon molecules or ``PAHs'' and is generally associated with the smallest interstellar grains.  The [CII] line is the dominant gas coolant in most regions of atomic interstellar gas, and therefore reflects the heating input to the gas.  The ratio of these two quantities, [CII]/AFE, remains nearly constant around 1.5\% against variations in the ratio of the \IRAS\ 60 $\mu$m band flux to the 100 $\mu$m band flux, R(60/100).  This is in contrast to the drop in the [CII]/FIR ratio with increasing R(60/100), which signals higher dust temperatures and more intense radiation fields.  We interpret the stable [CII]/AFE ratio as evidence that gas heating is dominated by the PAHs or small grains, which are also AFE carriers, over a wide range of conditions.  The trend of decreasing [CII]/FIR and AFE/FIR with increasing radiation field suggests a decrease in the importance of PAHs or small grains relative to large grains both in gas heating and in dust cooling.  We summarize the observed trends and suggest two plausible scenarios.

\end{abstract}
\altaffiltext{1}{IPAC, 100-22, California Institute of Technology, Pasadena, CA 91125} 
\altaffiltext{2}{Johns Hopkins University, Charles and 34th Street, Bloomberg center, Baltimore, MD 21210}
\altaffiltext{3}{Hubble Fellow} 
\altaffiltext{4}{NASA/Ames Research Center, MS 245-3, Moffett Field, CA 94035}


\section {Introduction}

The heating and cooling processes in the interstellar medium are crucial to
determining its physical state in equilibrium.  Photoelectrons from dust
grains or Polycyclic Aromatic Hydrocarbons (PAHs) are thought to dominate
the heating of neutral gas (Watson 1972; Hollenbach
\& Tielens 1999). In an indirect and inefficient mechanism, incident
far-ultraviolet photons with energies high enough to eject electrons from
dust grains ($h \nu \wig> 6$ eV) heat the gas via these photoelectrons,
with a typical efficiency of $0.1-1\%$.  This efficiency is determined by
microphysics of the grains, in particular the work function and the
resultant photoelectric yield, and therefore the charge of the grains.  It
is defined as the ratio of the gas heating rate by photoelectrons to grain
heating rate by far-ultraviolet photons.  The grain heating is radiated
away in the infrared.  The gas heating is also radiated away in the
infrared, primarily via fine-structure lines such as [CII] (158 $\mum$) and
[OI] (63 $\mum$), with [CII] dominant at lower densities and temperatures.
The photoelectric effect on dust thus provides the main coupling of the
far-ultraviolet radiation field to interstellar gas outside of \HII\
regions, and is a major process in photodissociation regions (Hollenbach \&
Tielens 1999).  The heating efficiency of the photoelectric effect is
traditionally measured by the flux ratio ([CII]+[OI])/FIR, where FIR is the
dust continuum in the far-infrared.

The physical description of the photoelectric effect and its dependence on
various parameters is fairly well developed.  In particular, it has been
established for some time that smaller grains are more effective at
generating photoelectrons, because of their lower potential barrier to
escaping electrons, and because the emerging electron is less likely to
lose energy as it works its way from the ejection site within the grain to
the outside.  Given the typical grain size distributions, the net result is
that the smallest grains dominate the photoelectric heating (Watson 1972,
Jura 1976, de Jong 1977, Draine 1978, Bakes \& Tielens 1994).  d'Hendecourt
\& L\'eger (1987) argued that PAHs may be the key to \HI\ interstellar gas
heating in clouds and intercloud regions, assuming they carry a plausible
10\% of interstellar carbon.  On the other hand, in a standard Bakes \&
Tielens model of the Orion Bar about half of the total photoelectric
heating is due to small grains with radii less than 15 \AA, typical
of PAHs and smaller molecules.  The importance of PAHs to the ionization
balance and chemistry of the interstellar medium is discussed by Lepp \&
Dalgarno (1988).

Quite independently, the hypothesis of very small grains heated
stochastically to emit at wavelengths shorter than expected for their
radiative environment was proposed by Andriesse (1978), and by Sellgren
(1984) to explain her data on reflection nebulae which showed an excess in
the near infrared with abnormally elevated apparent temperatures.  \IRAS\
revealed ubiquitous dust emission at 12 $\mum$, and very small grains were
accepted as widespread (Beichman 1987).  Combined with rudimentary spectral
data, the \IRAS\ 12 $\mum$ emission was associated with Aromatic Features
(Puget \& Leger 1989).  Detailed spectra with \ISO\ left no doubt that the
mid-infrared emission from dust in all but the most extreme environments is
dominated by the Aromatic Features, and is therefore a good tracer of PAHs
(Tielens et al. 1999).  Helou et al. (2000) recently
established that the mid-infrared spectra of normal galaxies are almost
constant in shape, and dominated by the Aromatic Features in Emission
(AFE), which carry from a few percent up to 20\% of the total infrared
luminosity.

\section {The [CII] ``Deficiency''}

Malhotra et al. (1997 and 2000a) reported \ISO\ observations of 60 normal
galaxies spanning a broad range in R(60/100), FIR/B-band flux and
morphology, and demonstrated that [CII]/FIR decreases as R(60/100)
increases in galaxies. Since most galaxies have ratios [CII]/FIR $\sim
10^{-3} - 10^{-2}$, those with ratios $<10^{-3}$ were said to be
``deficient.''  Since R(60/100) measures average temperatures of the large
dust grain population, it increases with increasing far-ultraviolet flux
$G_0$. The lowest [CII]/FIR ratios are observed in warmer and more actively
star-forming galaxies.  There have been many explanations proposed, such as
optical depth effects, heating radiation too soft to ionize carbon, or the
influence of an active galactic nucleus (Malhotra et al. 1997 and 2000a,
Luhman et al. 1998, Genzel \& C\'esarsky 2000).  The currently prevailing
opinion, however, points to decreased efficiency of photoelectric heating
due to positively charged grains. In the more actively star-forming and
warmer galaxies, the far-ultraviolet flux is higher. A higher ratio of
far-ultraviolet flux to gas density, $G_0/n$, leads to more positively
charged grains and therefore to lower photoelectric efficiency.

It is possible that the decrease in [CII]/FIR, instead of indicating a
decrease in the heating efficiency, might result from the gas cooling via
the other principal channel, [OI] 63 $\mu$m.  However, the combined flux in
the cooling lines normalized to the far-infrared emission ([CII]+[OI])/FIR
still shows a clear trend towards smaller values for the more actively
star-forming galaxies. [OI] 63 $\mum$ does become the dominant coolant for
galaxies showing warmer dust temperatures, but this relative rise does not
compensate for the decrease in [CII] relative to FIR.

\section {A New Normalization?}

The ``deficiency'' seen in [CII] is defined by comparison to the
far-infrared emission from dust grains.  Far-infrared emission measures the
radiation emitted by large or ``classical'' grains that are in thermal
equilibrium, maintaining a nearly constant temperature.  If photoelectric
heating is dominated by small grains, a more appropriate normalization
factor for [CII] line should be the emission from these grains (Helou 1999).
The smallest grains briefly reach very high effective temperatures with the
absorption of a single photon and therefore emit in the mid-infrared.  In
the PAH picture, this can be viewed as a fluorescence phenomenon from a
molecule at high excitation temperature.  In any case, because of the
invariant AFE spectrum, the mid-infrared flux tracks well the total heating
of this grain population.  Figure 1 (top panel) shows the [CII] line
strength normalized to the AFE flux plotted against R(60/100).  The AFE
flux is derived from ISO-CAM observations with broadband filter at 6.75
$\mum$ (Dale et al. 2000), which is very well matched to measuring the AFE.
The observed flux is scaled up by a factor of 1.17 to recover the
integrated 5--10 $\mum$ flux.  The scaling factor derives from the
properties of the ISO-CAM 6.75 $\mum$ filter and the average mid-infrared
spectrum of Helou et al. (2000).  We see not only a lack of any trend of
decreasing [CII]/AFE with R(60/100), but the scatter in the [CII]/AFE
values is about half the dispersion in [CII]/FIR (comparing top and
bottom panels of Figure 1). This suggests that there is indeed a special
connection between mid-infrared emission and [CII] emission.

\subsection{Scatter and Outliers}

The measured line fluxes have 30\% uncertainty which contributes
substantially to the observed scatter of $\sigma=0.18$ dex ($\simeq $
50\%).  The 5--10 $\mum$ flux attributed to AFE also includes some
contribution from stellar light which is not significant except in early
type galaxies (Malhotra et al. 2000b). The mid-infrared emission is from
grains heated by far-ultraviolet as well as optical photons (Uchida,
Sellgren \& Werner 1998), whereas only far-ultraviolet photons at energies
typically greater than 6 eV are effective in heating the gas.  Thus the
variation of the hardness of the radiation field also introduces some
scatter in the observed [CII]/AFE value. Some fraction of [CII] flux also
arises in ionized regions where the photoelectric effect is not the
dominant heating mechanism (Petuchowski \& Bennett 1993, Heiles 1994,
Malhotra et al. 2000a); extreme cases of this offset are discussed further
below.  With all these terms, it is remarkable that the dispersion in
[CII]/AFE is not greater than measured.

The lowest value of [CII]/AFE in Figure 1 represents NGC~4418 and is more
than 5$\sigma$ away from the mean. As discussed by Lu et al. (2000), the
mid-infrared spectrum of NGC~4418 is not typical of star-forming
galaxies. It shows none of the familiar aromatic features but instead has a
very broad plateau between 6 and 8.5 $\mum$. This may be a partial view of
the continuum along with the 9.7 $\mum$ silicate absorption feature. In any
case the imaging with the ISO-CAM filter at 6.75 $\mum$ does not measure
aromatic feature emission for NGC~4418. The upper limit in [CII]/AFE in
Figure 1 represents IC~860, from which [CII] was not detected.  The
mid-infrared spectrum of IC~860 appears similar to that of NGC~4418, but
has a very low signal-to-noise ratio (Lu et al. 2000).  Both galaxies are
OH mega-masers.

Two of the three highest values of [CII]/AFE in Figure 1 belong to galaxies
that have [NII]/FIR a factor of ten and 4$\sigma$ greater than the mean value.
As discussed by Malhotra et al. (2000a), this indicates a larger than average
contribution from \HII\ regions to the [CII] line emission from these
galaxies, which appears to cause somewhat elevated [CII]/AFE.

\subsection {[OI] $\lambda$ 63 $\mum$}

Since [OI](63 $\mum$) is also an important coolant, a more appropriate
quantity to plot might be the total gas cooling with respect to the AFE
flux, i.e.  ([CII]+[OI])/AFE.  In Figure 2 this quantity is plotted against
R(60/100), showing a larger scatter in the ([CII]+[OI])/AFE, but
essentially the same behavior as the ratio [CII]/AFE up to
R(60/100)$\sim0.65$.  For greater values, the scatter increases quickly as
does the mean ratio.  This is consistent with [CII] being the dominant
cooling line for all but the warmest galaxies (Malhotra et al. 2000a). The
energy level and critical density for [OI] (63 $\mum$) is higher than for
[CII], so this line is expected to become more important relative to [CII]
in warmer and dense gas, which is associated with greater values of $G_0/n$
and R(60/100) in photodissociation region models.

\section {Discussion}

The empirical results above strongly suggest that the smallest particles
which produce the mid-infrared emission are a key contributor to the
photoelectric effect in photodissociation regions.  Theoretical studies of
the Orion photodissociaton region where the values of $G_0$ and $n$ are
elevated and the ratio $G_0/n$ is also moderately high, show that about
half the photoelectric heating is due to particles smaller than 15 \AA\
(Bakes \& Tielens 1994).  The observed constancy of the [CII]/AFE ratio, in
a sample where the [CII]/FIR ratio shows larger systematic variations,
indicates that the role of small grains in heating the gas may be even more
prominent.  The interpretation of this empirical picture however is
hindered by serious uncertainties in the properties of the systems
involved, as detailed below.  We present below (\S \ref{sec:scenarios}) two
possible scenarios to account for the empirical facts, without favoring
either.  The empirical evidence can be summarized in four items:

\begin{description}
\item{[CII]/AFE has a small dispersion for R(60/100)$<0.6$, 
$\sigma\sim 0.16$ dex.}

\item{[CII]/FIR has a moderate dispersion for R(60/100)$<0.6$, 
$\sigma\sim 0.18$ dex, 
and falls off by an order of magnitude at
larger R(60/100) values (Malhotra et al. 2000a).}

\item{[OI]/FIR has a larger dispersion for R(60/100)$<0.6$,
$\sigma\sim0.20$ dex, and shows no convincing evidence of falling with
R(60/100); there are few data points at larger R(60/100) values, but the
dispersion appears to increase quickly.}

\item{[OI]/[CII] rises with R(60/100) (Malhotra et al. 2000a).}
\end{description}

\subsection {Uncertainties}

The exact composition of grains responsible for the spectral features in
the mid-infrared is uncertain, but these features correspond to bending and
stretching modes of C-C and C-H bonds in Aromatic hydrocarbons (Duley \&
Williams 1981).  The AFE have often been attributed to PAHs (L\'eger \&
Puget 1984; Allamandola, Tielens \& Barker 1985); we will adopt here the
notation PAH to refer to the carriers of the AFE.  It is also uncertain in
what proportion ionized and neutral PAH might contribute to the emission,
especially in light of recent \ISO\ observations that show little change in
mid-infrared spectral shapes as the far-ultraviolet flux ranges over
$G_0=1-10^5$ (Uchida et al. 2000).  These data appear to contradict
laboratory studies which show significant changes in relative strengths of
different features from ionized to neutral PAHs, with the ionized PAHs much
stronger in the 5 to 9 $\mum$ range.

In view of these uncertainties, we cannot unambiguously associate AFE with
a specific stage of the photoelectric process, even though the evidence in
Figure 1 suggests that the AFE closely track the rate of photoelectron
generation into the interstellar medium.  For instance, Figure 1 might be
easily understood if AFE derive predominantly from PAHs that have just been
ionized and are decaying from an excited state, or predominantly from
recombining PAHs, or from both in a roughly constant proportion.  However,
no such association can be established, and moreover excitation by
non-ionizing photons cannot be ruled out (Uchida et al. 2000).
Furthermore, there is currently no way to ascertain or constrain the
ionization balance of the carrier population.

Additional uncertainty arises from the significant variations known to
occur in the concentration of AFE carriers relative to large grains.
Boulanger et al. (1990) studied the relative distribution of 12 and
100~$\mum$ \IRAS\ emission in molecular clouds and concluded that the
variations of R(12/100) required the abundance of transiently heated
grains, which are mostly PAHs, to vary relative to the large grain
population.  In particular, they observed a systematic decrease of
R(12/100) as the 100~$\mum$ brightness increased, which appears as limb
brightening of clouds in AFE with a greater contrast than explained by
optical depth effects.  They proposed that this abundance variation is
related to the cycling of interstellar matter between gas phase and grain
surfaces.  On the other hand, the AFE contribution to the total dust
luminosity from galaxies decreases as the heating increases (Helou, Ryter
\& Soifer 1992; Lu et al. 2000).  This trend has been related to the
destruction of AFE carriers in the most intense heating environments near
and in \HII\ regions, a claim that is supported by observations of the
Milky Way and the Small Magellanic Cloud (Tran 1999, Contursi et al. 2000),
but may be linked to the effect observed by Boulanger et al. (1990).  These
depletion trends must then be superposed on the trends of photoelectric
yield with interstellar medium parameters, resulting in the empirical
evidence summarized above and in Figure 1.

\subsection {Possible Scenarios}
\label{sec:scenarios}
  
The close association of [CII] with AFE is easy to interpret at low values
of R(60/100), by assuming that [CII] represents the cooling of the gas, and
AFE measures the heating by photoelectrons.  However, at larger values of
R(60/100), reflecting an enhanced $G_0/n$ (Malhotra et al. 2000a), this
association becomes puzzling since [OI] overtakes [CII] in cooling
luminosity, casting doubt on the argument that [CII]/AFE is constant
because it is intimately tied to the photoelectric effect.  As the heating
becomes more intense, the ratio of ([CII]+[OI])/AFE rises, suggesting an
apparent increase in the photoelectric efficiency of the PAHs; this however
is in direct conflict with the physics, leading us to the conclusion that
PAHs become less significant contributors to the total photoelectric
heating compared with other grain populations.  This decreased significance
is consistent with the decreased energetic importance of PAHs, manifested
as the drop in AFE/FIR with increasing R(60/100) (Helou, Ryter \& Soifer
1991).

In this picture of decreasing significance of [CII] as a cooling line and
of AFE as a dust radiator, one would have to invoke a coincidental
agreement between these two rates of decrease in order to explain the
unchanged ratio of [CII]/AFE.  This coincidence may reflect a certain
scaling of $G_0$ with $n$, similar to the $G_0\propto{n}^{1.4}$ derived by
Malhotra et al. (2000a).  Such a scaling might result from a feedback
mechanism linking the density of the medium and the intensity of the star
formation, turning the coincidence into a causal connection.  In any case,
it is not clear what drives the systematic decrease in AFE/FIR, though it
has been suggested that a more intense radiation field may be inducing
greater ionization or de-hydrogenation of the PAHs (Tielens et al. 1999).

An alternate scenario to explain the observations would be a two-component
model broadly similar to the Helou (1986) decomposition of the far-infrared
emission into an ``active'' star formation component and a ``quiescent'' or
cirrus component.  The cirrus-like component would dominate at low values
of R(60/100), and therefore be characterized by high values of [CII]/[OI],
AFE/FIR and [CII]/FIR, just as one would expect from a photodissociation
region with low density and $G_0/n$.  On the other hand, the high-density
photodissociation regions at the surface of molecular clouds directly
illuminated by young massive stars would provide the active component, with
elevated values of $G_0, n$ and $G_0/n$, resulting in more cooling via [OI]
than [CII], and less efficient gas heating due to grain charging.  In this
scenario, one would have to assume that the AFE emission from the active
component is depressed compared to the diffuse component, a property
marginally suggested by the observations of lower AFE/FIR ratios in denser
molecular clouds (Boulanger et al. 1996).  A significant contribution to
the active component by \HII\ regions would also help explain the depressed
values of AFE/FIR.  The observed trends as a function of R(60/100) are then
interpreted as a result of a varying proportion between the two components,
with the interstellar medium falling almost bimodally into these two
components, and the properties of each component varying little among
galaxies.

While either one of these scenarios is plausible, each requires a number of
assumptions which are difficult to verify with currently available data.
This however illustrates the richness of the data at hand, and the need for
a more sophisticated treatment of the interstellar medium on the scale of
galaxies.

\acknowledgements 
We would like to thank F. Boulanger \& J.-L. Puget for stimulating discussions and J. Brauher for help with data processing.  This work was supported by \ISO\ data analysis
funding from NASA, and carried out at IPAC/JPL of the California Institute of Technology.  \ISO\ is an ESA project with instruments funded by ESA Member States (especially the PI countries: France, Germany, the Netherlands and the United Kingdom), and with the participation of ISAS and NASA.  SM's research funding is provided by NASA through Hubble Fellowship grant \#HF-01111.01-98A from the Space Telescope Science Institute, which is operated by AURA under NASA contract NAS5-26555. This research has made use of the NASA/IPAC Extragalactic Database which is operated by the Jet Propulsion Laboratory, California Institute of Technology, under contract with NASA.

\begin{figure}[ht]
\plotone{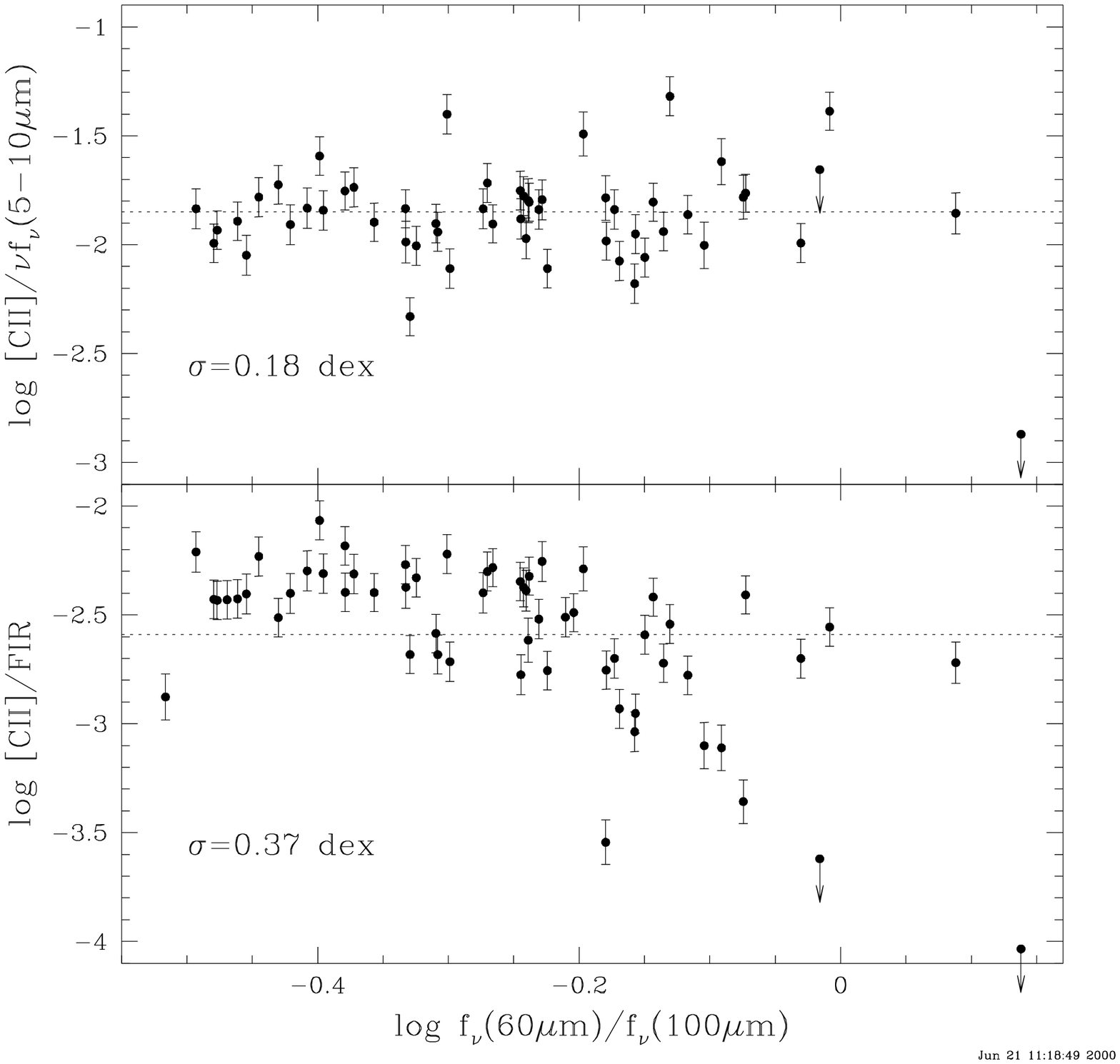}
\caption{ The ratio [CII]/AFE is seen to be constant with FIR colors
R(60/100) (top panel). This is in contrast to the ratio [CII]/FIR which
declines for galaxies with higher R(60/100) colors, i.e. warmer dust
(bottom panel). The constancy of [CII]/AFE suggests that the small grains
which are responsible for mid-infrared emission also dominate the
photoelectric heating of gas.  The 1$\sigma$ scatter is with respect to the
mean ratio, which is indicated by the dotted lines in each panel.}
\end{figure}

\begin{figure}[ht]
\plotone{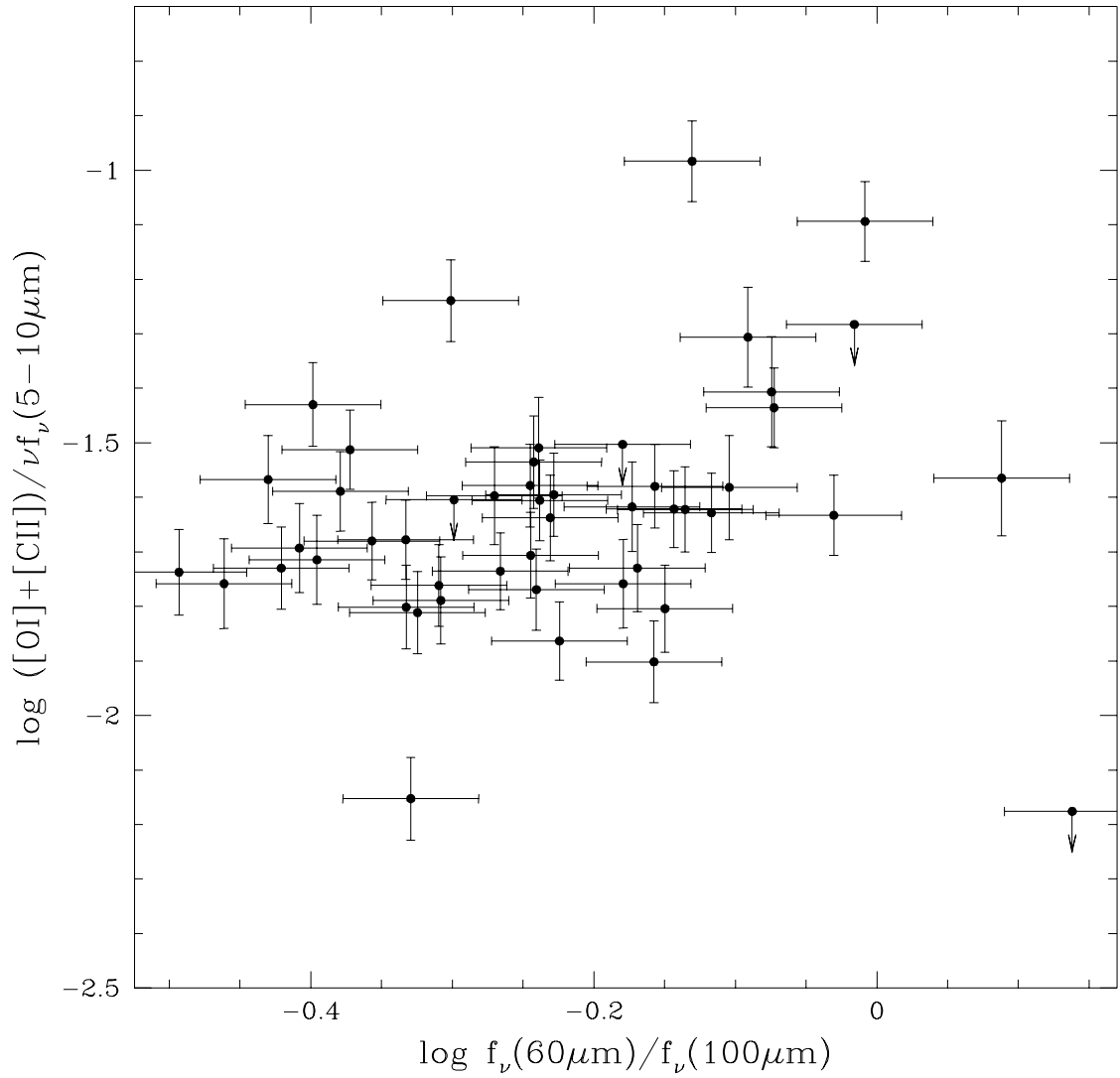}
\caption{ The ratio ([CII]+[OI])/AFE is plotted against R(60/100). 
The scatter in the ([CII]+[OI])/AFE is larger than that for [CII]/AFE.}
\end{figure}

\end{document}